\documentclass[twocolumn,showpacs,preprintnumbers,amsmath,amssymb]{revtex4}
\usepackage{rotating}
\usepackage{bm}% bold math
\usepackage[colorlinks=true,linkcolor=blue,citecolor=blue,urlcolor=blue]{hyperref}% add hypertext capabilities
\usepackage{color}
\usepackage{array,multirow,graphicx}
\usepackage{epstopdf}

\begin{document}

%\preprint{APS/123-QED}
\newcommand{\quotes}[1]{`#1'}
\title{The 10\% Gd and Ti co-doped BiFeO$_3$: A promising multiferroic material}

\author{M. A. Basith}
\email[Author to whom correspondence should be addressed (e-mail): ]{mabasith@phy.buet.ac.bd}
\author{Areef Billah}
\author{M. A. Jalil}
\author{Nilufar Yesmin}
%\altaffiliation[Present address: ]{Department of Physics, Bangladesh University of Engineering and Technology, Dhaka-1000, Bangladesh.}
\affiliation{Department of Physics, Bangladesh University of Engineering and Technology, Dhaka-1000, Bangladesh.
}
\author{Mashnoon Alam Sakib, Emran Khan Ashik, S. M. Enamul Hoque Yousuf, and Sayeed Shafayet Chowdhury}
%\altaffiliation[Present address: ]{Department of Physics, Bangladesh University of Engineering and Technology, Dhaka-1000, Bangladesh.}
\affiliation{Department of Electrical and Electronic Engineering, Bangladesh University of Engineering and Technology, Dhaka-1000, Bangladesh.
}
\author{Md. Sarowar Hossain }
%\email[Author to whom correspondence should be addressed (e-mail): ]{mabasith@phy.buet.ac.bd}
%\altaffiliation[Present address: ]{Department of Physics, Bangladesh University of Engineering and Technology, Dhaka-1000, Bangladesh.}
\affiliation{S. N. Bose National Centre for Basic Sciences, Salt Lake City, Kolkata, West Bengal 700098, India.
}	
\author{Shakhawat H. Firoz}
\affiliation{Department of Chemistry, Bangladesh University of Engineering and Technology, Dhaka-1000, Bangladesh.
}
\author{Bashir Ahmmad}
\email[Author to whom correspondence should be addressed (e-mail): ]{arima@yz.yamagata-u.ac.jp}
\affiliation{Graduate School of Science and Engineering, Yamagata University, 4-3-16 Jonan, Yonezawa 992-8510, Japan.}

\date{\today}% It is always \today, today,
             %  but any date may be explicitly specified

\begin{abstract}
 In this investigation, undoped BiFeO$_3$, Gd doped Bi$_{0.9}$Gd$_{0.1}$FeO$_3$, and Gd-Ti co-doped Bi$_{0.9}$Gd$_{0.1}$Fe$_{1-x}$Ti$_x$O$_3$ (x = 0.10, 0.20) materials were synthesized to report their multiferroic properties. The structural analysis and phase identification of these multiferroic ceramics were performed using Rietveld refinement. The Rietveld analysis has confirmed the high phase purity of the 10\% Gd-Ti co-doped Bi$_{0.9}$Gd$_{0.1}$Fe$_{0.9}$Ti$_{0.1}$O$_3$ sample compared to that of other compositions under investigation. The major phase of this particular composition is of rhombohedral \textit{R3c} type structure (wt\% $>99\%$) with negligible amount of impurity phases. In terms of characterization, we address magnetic properties of this co-doped ceramic system by applying substantially higher magnetic fields than that applied in previously reported investigations. The dependence of temperature and maximum applied magnetic fields on their magnetization behavior have also been investigated. Additionally, the leakage current density has been measured to explore its effect on the ferroelectric properties of this multiferroic system. The outcome of this investigation suggests that the substitution of 10\% Gd and Ti in place of Bi and Fe, respectively, in BiFeO$_3$ significantly enhances its multiferroic properties. The improved properties of this specific composition is associated with homogeneous reduced grain size, significant suppression of impurity phases and reduction in leakage current density which is further asserted by polarization vs. electric field hysteresis loop measurements.

 %The manuscript presented is an extension of our previous research on room temperature dielectric and magnetic properties of Gd and Ti co-doped BiFeO$_3$ ceramics (J. App. Phys. 115, 024102, 2014). In this investigation, 
\end{abstract}

%\keywords{Suggested keywords}%Use showkeys class option if keyword
                              %display desired
\maketitle
\section{Introduction} \label{I}
Recently, there has been a great interest for the study of the multiferroic materials, in which ferromagnetic, ferroelectric, and/or ferroelastic orderings coexist \cite{ref61,ref1,ref2,ref3, ref62}. The co-existence of \quotes{ferro}-orders in multiferroics opens up pathways for the possibility that the magnetization can be controlled by the electric field and vice versa. The ability to manipulate the magnetic and ferroelectric properties of multiferroic BiFeO$_3$ (BFO) by dopants opens up promising opportunities for fabricating new multiferroic materials in the field of information storage technology. Noticeably, the spiral modulated spin structure (SMSS) of BFO possesses an incommensurate long-wavelength period of 62 nm \cite{ref63} due to which the macroscopic magnetization gets cancelled. As a result, the linear magnetoelectric effect is no longer observed \cite{ref64}. Moreover, the preparation of undoped BFO is challenging due to the formation of different impurity phases \cite{ref78, ref79}. Thus the use of undoped bulk BFO in functional applications gets rather limited due to these hindrances.

Recent investigations have demonstrated that simultaneous minor substitution of Bi and Fe in  BiFeO$_3$  by ions such  as La and Mn, La and Ti, Nd and Sc, Gd and Mn etc., respectively \cite{ref71, ref72, ref74, ref901} enhanced the magnetism and ferroelectricity in BiFeO$_3$. Lately, in a particular case, improvements in the morphological, dielectric and magnetic properties of BiFeO$_3$ multiferroics have been obtained at room temperature (RT) by simultaneous substitution of Gd and Ti in place of Bi and Fe, respectively in BiFeO$_3$ \cite{ref6}. Notably, Gd and Ti have been chosen to substitute at A-site and B-site, respectively of the ABO$_3$ (A = Bi, B = Fe) structure. At site A, Gd was chosen since previously, it was observed that the substitution of 10\% Gd in place of Bi resulted in enhancement of room temperature magnetization \cite{ref701, ref702} as well as improvement in phase purity of bulk BiFeO$_3$. Again, at B site, the partial substitution of Fe by Ti was performed as the substitution of Ti$^{4+}$ was reported to decrease the leakage current significantly and induce a remanent magnetization in BiFeO$_3$ \cite{ref703, ref704}. Moreover, the bond enthalpies of Gd-O ($719\pm 10$ kJ/mol) and Ti-O ($672.4\pm 2$ kJ/mol) bonds are, respectively, stronger than those of Bi-O ($337\pm 12.6$ kJ/mol) and Fe-O ($390\pm 17.2$ kJ/mol) bond enthalpies \cite{ref705}. This justifies the co-substitution of Gd and Ti, in place of Bi and Fe respectively, as we anticipate this would potentially recover the oxygen vacancies caused by volatilization of Bi atoms.

We have also observed a strong influence of temperature on coercive fields and exchange bias fields of this Gd and Ti co-doped  BiFeO$_3$  ceramic system \cite{ref101}. The coercive fields of this multiferroic system enhanced anomalously with increasing temperature \cite{ref101}. Our previous investigations also demonstrated that preparation of the nanoparticles of this co-doped system with size less than or around 62 nm of SMSS further improved their structural and magnetic properties \cite{ref102, ref103}. Later on, other research groups also observed interesting magnetic properties in this Gd and Ti co-doped BiFeO$_3$ ceramic system at RT \cite{ref73, ref104, ref601}. In Refs. \cite{ref6, ref73, ref104, ref601}, the magnetic properties were investigated by applying magnetic fields ranging from 15-20 kOe without measuring the dependence of magnetization on temperature. Moreover, up to date, there is no comprehensive result that have been reported on the multiferroicity of this co-doped material system. Therefore, in this investigation, we have carried out field dependent magnetic measurements of this ceramic system by applying magnetic fields of up to 50 kOe. The dependence of magnetization parameters on different applied magnetic fields ranging from 10 kOe to 70 kOe was also studied. Moreover, the effects of temperature on these ceramic materials were investigated as well. Along with magnetic characterization, the leakage current behavior and their effect on the ferroelectric properties were also part of our investigation. The compendium of these investigations is that the nominal composition of 10\% Gd and Ti co-doped BiFe0$_3$ having significantly improved phase purity exhibited enhanced multiferroic properties.  

%Finally, the temperature dependent dielectric properties of a selected sample were measured to confirm the presence of magnetoelectric coupling in this multiferroic material system.

\section{Experimental details} \label{II} 
The polycrystalline samples with compositions of undoped  BiFeO$_3$ (referred to as undoped BFO), Gd doped Bi$_{0.9}$Gd$_{0.1}$FeO$_3$ (referred to as Gd doped BFO) and Gd-Ti co-doped Bi$_{0.9}$Gd$_{0.1}$Fe$_{1-x}$Ti$_x$O$_3$ (x = 0.10-0.20) (referred to as Gd-Ti co-doped BFO) were synthesized by using standard solid state reaction technique as delineated in our previous investigation \cite{ref6}. To synthesize the polycrystalline samples for our investigation, the compacted mixtures were calcined at 800$^o$C for 1.5 hours in a programmable furnace. The calcined powders were grounded again for 2 hours to get more homogeneous mixture. The powders were then pressed into pellets using a uniaxial hydraulic press and sintered at 825$^o$C for 5 hours at heating rate of 10$^o$C per minute. The crystalline structures of these materials in bulk powder form were determined from X-ray diffraction (XRD) data using a diffractometer (Rigaku Smart Lab) with CuK$_{\alpha}$  (${\lambda}$ = 1.5418 $\AA$) radiation. X-ray photoelectron spectroscopy (XPS, ULVAC-PHI Inc., Model 1600) analysis was carried out with a Mg-K${\alpha}$ radiation source. The $M-H$ hysteresis loops of undoped BFO, Gd doped BFO and Gd-Ti co-doped BFO multiferroic ceramics were investigated using a Superconducting Quantum Interference Device (SQUID) Magnetometer (Quantum Design MPMS-XL7, USA). The temperature dependent magnetization measurements were investigated both at zero field cooling (ZFC) and field cooling (FC) processes \cite{ref103}. The leakage current density and ferroelectric polarization of the pellet shaped samples were traced using a ferroelectric loop tracer in conjunction with an external amplifier (10 kV). %The temperature dependent dielectric properties were measured using an impedance analyzer (Wayne Kerr) at different frequencies. 

\section{Results and discussion} \label{III}
\subsection{Structural Characterization} \label{I}
The XRD  patterns of undoped BFO, Gd doped BFO and Gd-Ti co-doped BFO ceramics, sintered  at temperature of their optimum density indicate the formation of polycrystalline structure. The formation of secondary phases during the solid state synthesis of bulk undoped BFO and cations substituted BFO was almost unavoidable in a number of previous investigations \cite{ref401, ref402, ref403, ref404}. The final distribution of these secondary phases \cite{ref405} was influenced greatly by the specific reaction pathway, however, the latent mechanism behind this formation is still unknown.
\begin{figure}[hh]
	\centering
	\includegraphics[width=9.5cm,height=13cm]{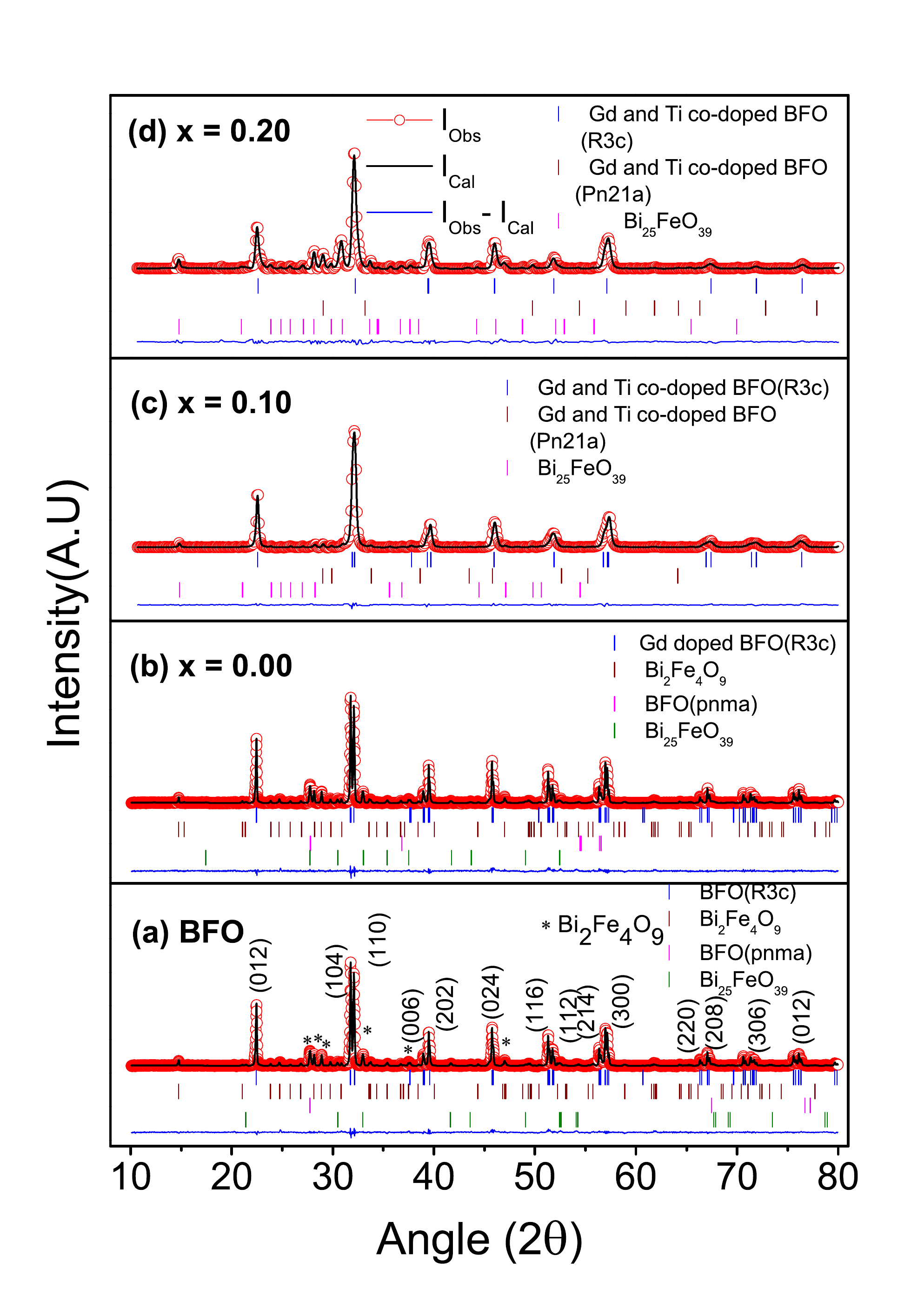}
	\caption{Rietveld refined XRD patterns of (a) undoped BiFeO$_3$, (b) Gd doped Bi$_{0.9}$Gd$_{0.1}$Fe$_{1-x}$Ti$_x$O$_3$ (x = 0.00), (c) Gd-Ti co-doped Bi$_{0.9}$Gd$_{0.1}$Fe$_{1-x}$Ti$_x$O$_3$ (x = 0.10) and (d) Gd-Ti co-doped Bi$_{0.9}$Gd$_{0.1}$Fe$_{1-x}$Ti$_x$O$_3$ (x = 0.20) ceramics carried out at RT. Here the observed data are represented by red circles ($Y_{obs}$), the black solid line represents the calculated pattern ($Y_{cal}$) and the blue bottom curve is their difference ($Y_{obs}$-$Y_{cal}$). } \label{fig01}
\end{figure}

%green bars correspond to Bragg positions of the major phase of the corresponding samples.

In our previous investigation \cite{ref6}, the structural analysis and phase identification of these multiferroic ceramics were not performed using Rietveld refinement. This time around, Rietveld refinement was carried out using the FULLPROF package \cite{ref6n} to analyze the crystal structure as well as to quantify the crystallographic phases of these compounds. The Rietveld refined XRD patterns of the four samples under scrutiny are depicted in Fig. \ref{fig01}. The structural parameters obtained with the help of the refinement as well as phases present (in wt\%) obtained from XRD studies of BFO, Gd doped BFO and Gd-Ti co-doped BFO are listed in supplemental Table 1 \cite{refst}. In each of the samples, we found that the major phases namely, BFO, Gd doped BFO and Gd-Ti co-doped BFO are of rhombohedral \textit{R3c} type crystal structure. However, the substitution of Bi by Gd indeed destabilizes the \textit{R3c} phase of BiFeO$_3$ \cite{catscott}. Hence, in BFO, besides the major phase, the impurity phases Bi$_2$Fe$_4$O$_9$ ($1.62\%$) and Bi$_{25}$FeO$_{39}$ ($0.10\%$) are present as well as shown in Fig.\ref{fig01}. Next, for the Gd doped BFO, we observed the presence of the same impurity phases. Interestingly, when we performed Gd and Ti co-substitution, the unintended phase is suppressed significantly as the wt\% of Bi$_2$Fe$_4$O$_9$ reduces to $0.39\%$ in 10\% Gd and Ti co-doped Bi$_{0.9}$Gd$_{0.1}$Fe$_{0.9}$Ti$_{0.1}$O$_3$. Furthermore, when the Ti doping is increased to 20\% in Bi$_{0.9}$Gd$_{0.1}$Fe$_{0.8}$Ti$_{0.2}$O$_3$, the amount of undesired Bi$_2$Fe$_4$O$_9$ phase slightly increases to $~ 0.95\%$ which substantiates our idea of the 10\% Gd and Ti co-doped BFO, Bi$_{0.9}$Gd$_{0.1}$Fe$_{0.9}$Ti$_{0.1}$O$_3$ being the most suitable composition to perform extensive investigation of its multiferroic properties. This is also illustrated in Fig. \ref{fig01} from the difference between the observed and calculated XRD patterns of the samples obtained from Rietveld refinement. We note that highest match occurs between these two patterns for 10\% Gd and Ti co-doped Bi$_{0.9}$Gd$_{0.1}$Fe$_{0.9}$Ti$_{0.1}$O$_3$ which indicates very negligible impurities present in this particular composition. The atomic coordinates of the samples have also been listed in the supplemental Table 1 \cite{refst}. Additionally, the errors in estimated structural parameters were included.

\begin{figure}[hh]
	\centering
	\includegraphics[width=9.5cm]{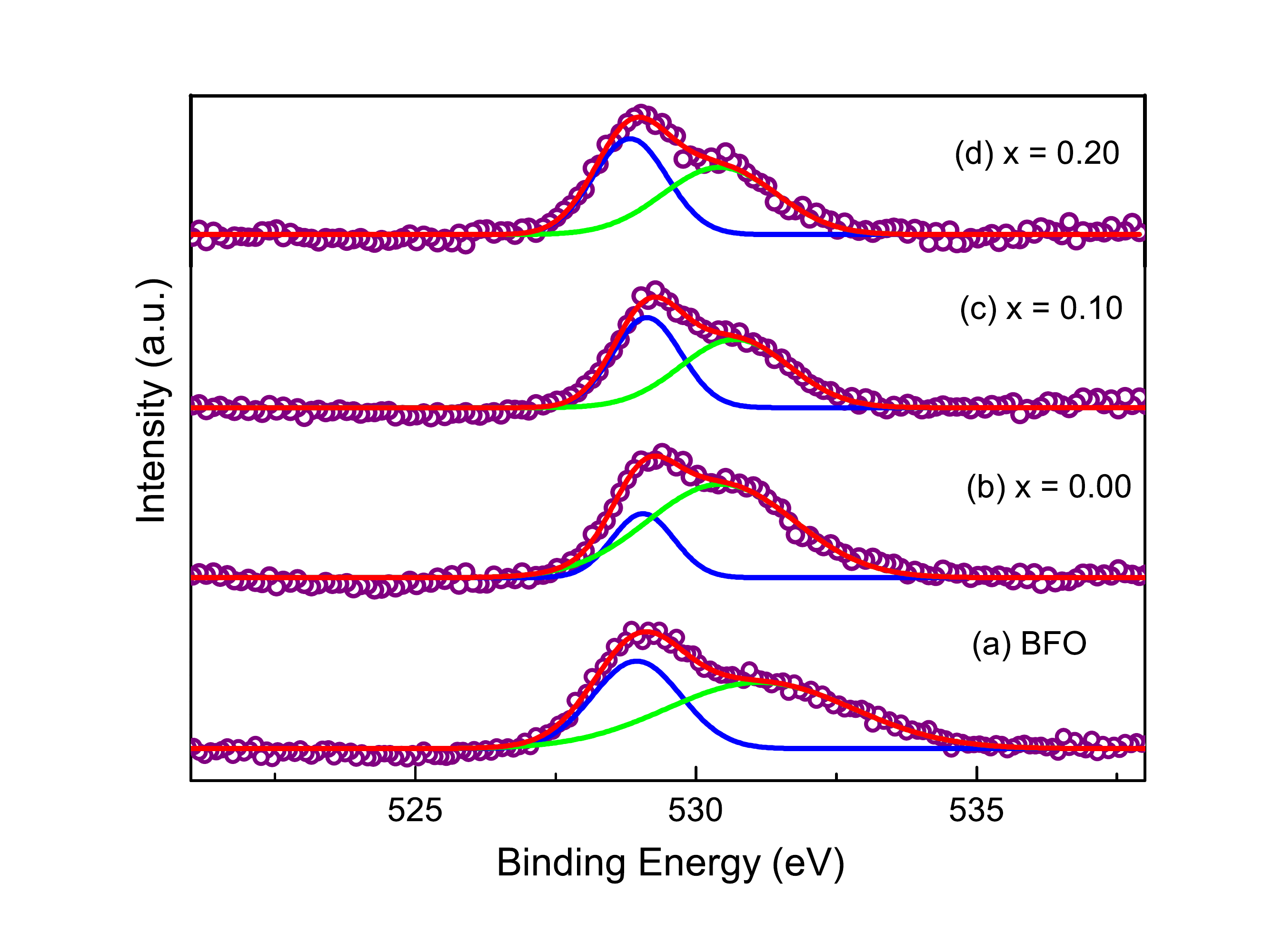}
	\caption{The O 1s core XPS spectrum of (a) undoped BiFeO$_3$, (b) Gd doped Bi$_{0.9}$Gd$_{0.1}$Fe$_{1-x}$Ti$_x$O$_3$ (x = 0.00), (c) Gd-Ti co-doped Bi$_{0.9}$Gd$_{0.1}$Fe$_{1-x}$Ti$_x$O$_3$ (x = 0.10) and (d) Gd-Ti co-doped Bi$_{0.9}$Gd$_{0.1}$Fe$_{1-x}$Ti$_x$O$_3$ (x = 0.20) samples.} \label{fig02}
\end{figure}

The X-ray photoelectron spectroscopy (XPS) of the above mentioned samples was performed. To investigate the oxygen vacancy related effects, in Fig. \ref{fig02} the O 1s core XPS spectra of the corresponding samples are shown. The plots display an asymmetric peak very close to 529 eV along with an additional peak at slightly higher binding energy (HBE). The asymmetric curves of the samples have been Gaussian fitted by two symmetrical peaks. The lower binding energy (LBE) peak around 529.3 eV corresponds to the O 1s core spectrum, while the HBE peak is related to the oxygen vacancy in the samples \cite{ref303n}. The area ratios of the two peaks (HBE/ LBE) for the samples BFO, Gd doped BFO, Bi$_{0.9}$Gd$_{0.1}$Fe$_{0.9}$Ti$_{0.1}$O$_3$ and Bi$_{0.9}$Gd$_{0.1}$Fe$_{0.8}$Ti$_{0.2}$O$_3$ are 1.59, 3.46, 1.21 and 1.07 respectively. So, clearly the ratio decreases with Gd-Ti co-substitution as compared to BFO and Gd doped BFO indicating reduction in oxygen vacancies. The effect of this reduced vacancy is further explored in section C where we investigate the leakage current density of the samples.

\subsection{Magnetic Characterization} \label{I}
For magnetic characterization, the $M-H$ hysteresis loops of undoped, Gd doped and Gd-Ti co-doped BFO ceramics were carried out at RT with an applied magnetic field of up to $\pm$50 kOe. The undoped BFO sample possesses a very narrow hysteresis loop with a very small but non-zero remanent magnetization (0.001 emu/g) and a coercive field of $\sim {~}$132 Oe at RT. This is due to antiferromagnetic (AFM) nature of undoped BFO which possesses no spontaneous magnetization \cite{ref301} but has residual magnetic moment for a canted spin structure. The Gd doped Bi$_{0.9}$Gd$_{0.1}$FeO$_3$  and Gd-Ti co-doped Bi$_{0.9}$Gd$_{0.1}$Fe$_{1-x}$Ti$_x$O$_3$ (x = 0.10-0.20) samples also exhibit unsaturated hysteresis loops but with large remanent magnetizations and coercive fields. Due to the substitution of Gd and co-substitution of Gd and Ti in place of Bi and Fe in BiFeO$_3$, respectively, the unsaturated magnetization behavior of the samples remains unaltered, however, the center of the hysteresis loops was many folds wider compared to that of undoped BFO. 

\begin{figure}[hh]
	\centering
	\includegraphics[width=10cm]{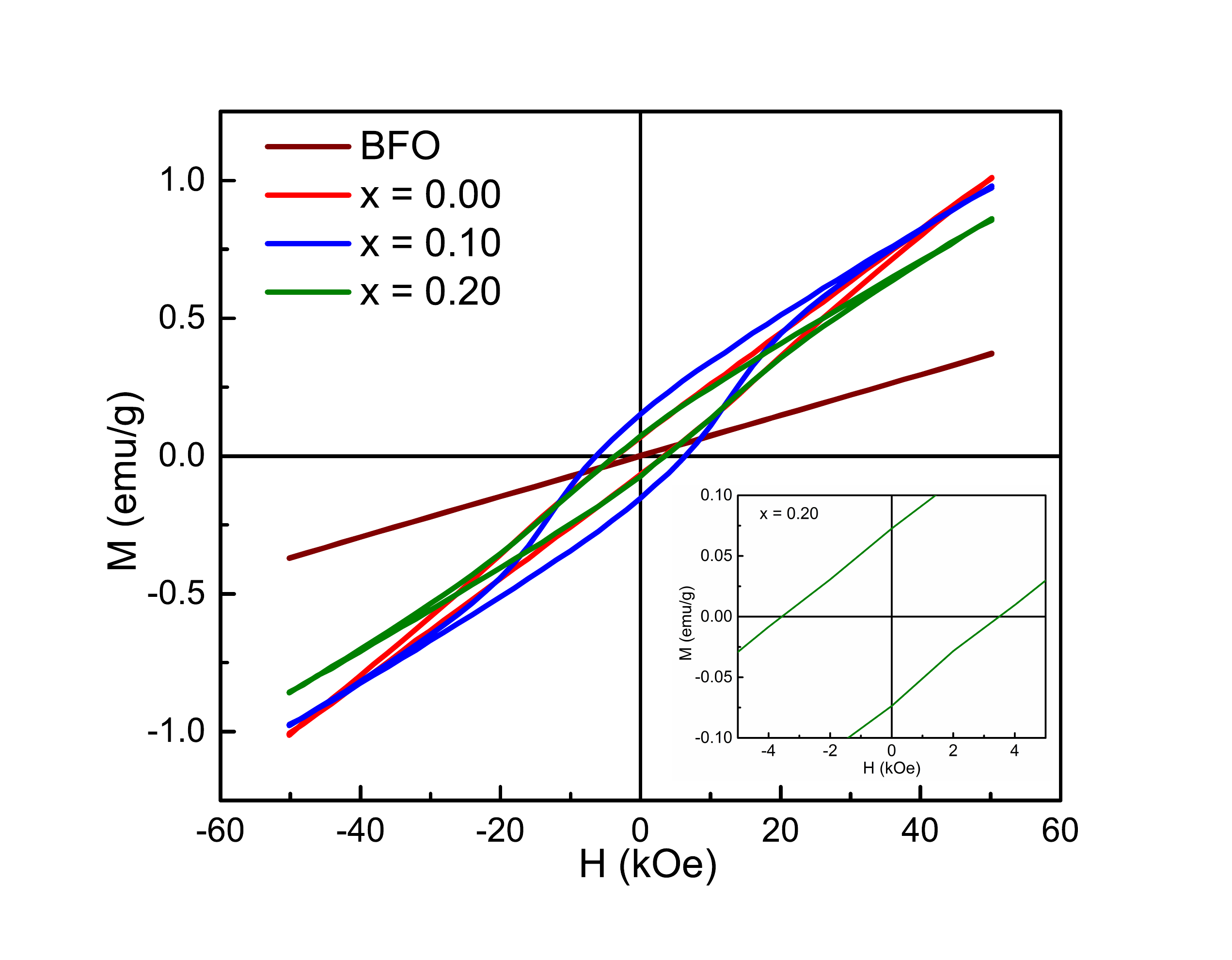}
	\caption{The $M-H$ hysteresis loops of undoped BiFeO$_3$, Gd doped Bi$_{0.9}$Gd$_{0.1}$Fe$_{1-x}$Ti$_x$O$_3$ (x = 0.00) and Gd-Ti co-doped Bi$_{0.9}$Gd$_{0.1}$Fe$_{1-x}$Ti$_x$O$_3$ (x = 0.10-0.20) ceramics carried out at RT with an applied magnetic field of up to $\pm$50 kOe. The inset shows the enlarged view of the loop for sample x = 0.20.} \label{fig1}
\end{figure}

For obtaining quantitative measures of the coercive fields (H$_c$) and remanent magnetization (M$_r$) from the hysteresis loops, formulas used were: $H_c = (H_{c1}-H_{c2})/2$, where H$_{c1}$ and H$_{c2}$ are the left and right coercive fields \cite{ref6, ref302} and M$_{r}$ = $|$(M$_{r1}$-M$_{r2}$)$|$/2 where M$_{r1}$ and M$_{r2}$ are the magnetization with positive and negative points of intersection with H = 0, respectively \cite{ref71}. Calculated values of M$_{r}$ and H$_c$ for undoped, Gd doped and Gd-Ti co-doped BFO bulk materials are inserted in Table \ref{Tab1}. Both the coercive fields and remanent magnetizations are higher for 10\% Gd and Ti co-doped Bi$_{0.9}$Gd$_{0.1}$Fe$_{1-x}$Ti$_{x}$O$_3$ (x = 0.10) sample than those for other materials. However, when we further increased the Ti concentration to 20\% in Bi$_{0.9}$Gd$_{0.1}$Fe$_{1-x}$Ti$_{x}$O$_3$ (x = 0.20), H$_c$ and M$_{r}$ got reduced despite their net values being still higher compared to that of Gd doped Bi$_{0.9}$Gd$_{0.1}$FeO$_3$ sample. The larger values of H$_c$ and M$_r$ in sample x = 0.10 may be associated with the microstructure of the composition, i.e. with its homogeneous reduced grain size than that of the other materials as reported in our previous investigation \cite{ref6}. To demonstrate the homogeneous reduced grain size of 10\% Gd and Ti co-doped BFO multiferroic material, FESEM micrographs were inserted in the supplementary information (supplementary figure S1). Notably,  Zhai \textit{et al.} attributed increase in coercive field with the co-substitution of La and Nb in BiFeO$_3$ \cite{ref303}. Similar reasoning was reported for Pr and Zr co-substituted BiFeO$_3$ compounds in Ref. \cite{ref304}. The unsaturated magnetization behaviour at higher fields clearly indicates the dominating AFM nature of these ceramics. A large coercive field, of up to 6399 Oe is observed for 10\% Gd and Ti co-doped Bi$_{0.9}$Gd$_{0.1}$Fe$_{0.9}$Ti$_{0.1}$O$_3$ sample owing to the strong magneto-crystalline anisotropy of the compound despite having a FM component.

\begin{table}[!h]
	\caption{The table shows the calculated values of M$_{r}$, H$_{c}$ and H$_{EB}$ for undoped BiFeO$_3$, Gd doped Bi$_{0.9}$Gd$_{0.1}$Fe$_{1-x}$Ti$_x$O$_3$ (x = 0.00) and Gd-Ti co-doped Bi$_{0.9}$Gd$_{0.1}$Fe$_{1-x}$Ti$_x$O$_3$ (x = 0.10-0.20) ceramics observed at RT.}
	\label{Tab1}   
	\begin{center}
		\begin{tabular}{|l|l|l|l|} 
			\hline
			\cline{1-4}
			Samples&M$_{r}$ (emu/g)&H$_{c}$ (Oe)&H$_{EB}$ (Oe)\\

			\hline
			BFO&0.001&132&81\\
			\hline
			0.00&0.065&3406&33\\
			\hline
			0.10&0.155&6399&-13\\
			\hline
			0.20&0.073&3532&40\\
			\hline
		\end{tabular}
	\end{center}
\end{table}

An asymmetric shift towards the magnetic field axes \cite{ref103} in the $M-H$ hysteresis loops at RT is observed as depicted in Fig.~\ref{fig1}. The asymmetry was shown in the inset of Fig. \ref{fig1} for sample x = 0.20. The presence of an exchange bias (EB) effect is evinced by this asymmetry phenomenon in this multiferroic material \cite{ref6, ref103, ref305}. As mentioned earlier, the hysteresis loops of these ceramics showed in Fig. \ref{fig1}, confirm the basic AFM nature of the compounds. We notice that the centers of $M-H$ loops of Bi$_{0.9}$Gd$_{0.1}$Fe$_{1-x}$Ti$_x$O$_3$ (x = 0.00-0.20) compounds get widened when compared to undoped BFO. This provides an indication of their weak ferromagnetic nature \cite{ref306}. The temperature dependent magnetization curves \cite{ref306} further assert this weak ferromagnetic nature as described later on. We presume that this multiferroic material system bears the coexistence of the anisotropic ferri/ferromagnetic (FM) and anti-ferromagnetic domains. The exchange coupling at the interfaces between the multiple magnetic domains, forces the system to act as a natural system for generating EB effect \cite{ref302, ref307, ref308, ref309}.

\begin{figure}[hh]
	\centering
	\includegraphics[width=10cm]{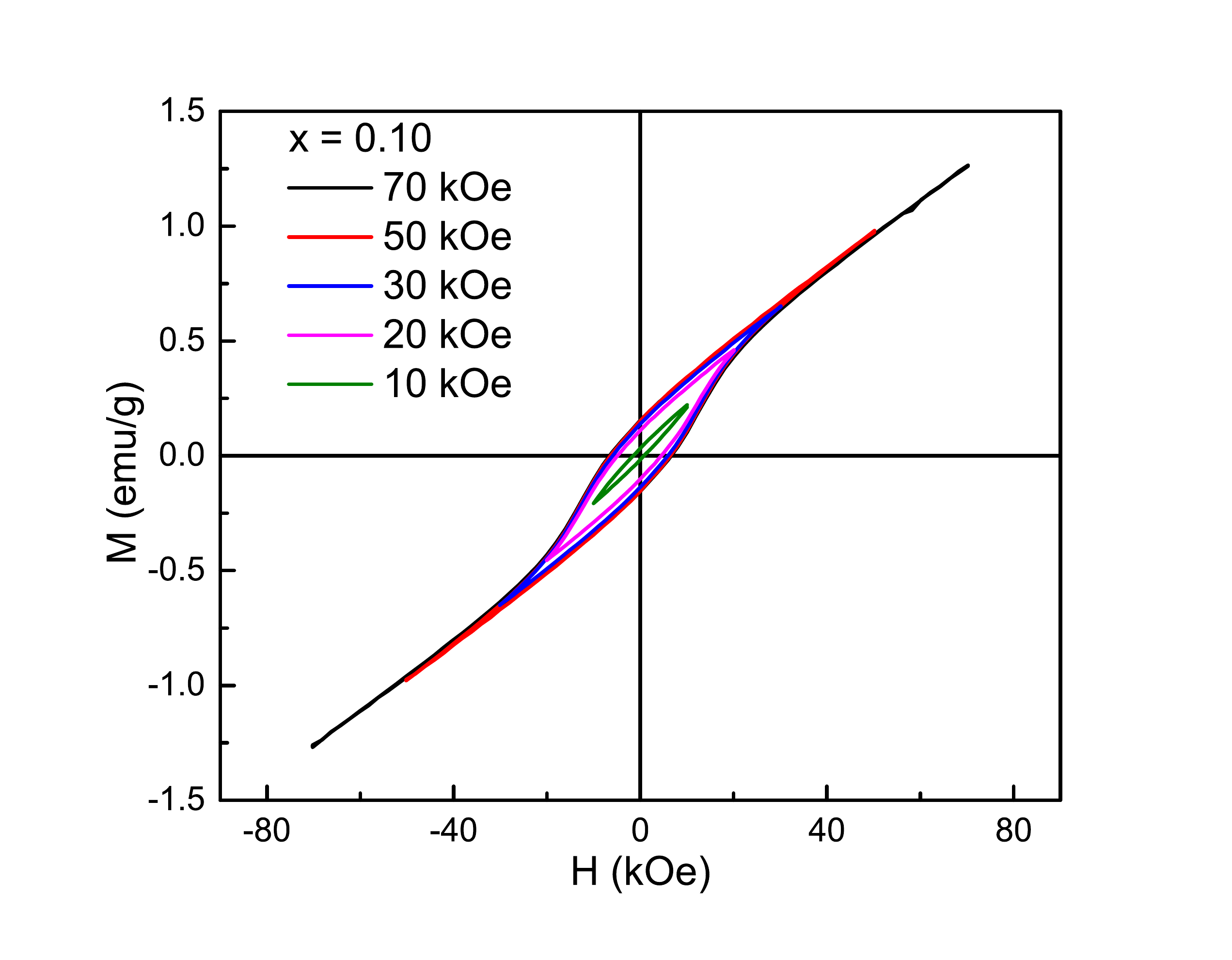}
	\caption{The $M-H$ hysteresis loops of 10\% Gd and Ti co-doped Bi$_{0.9}$Gd$_{0.1}$Fe$_{0.9}$Ti$_{0.1}$O$_3$ ceramics for different H$_{max}$ at RT.} \label{fig2}
\end{figure}

The quantification of the exchange bias field (H$_{EB}$) found from the loop asymmetry along the field axis was performed using $H_{EB} = -(H_{c1}+H_{c2})/2$ where $H_{c1}$ and $H_{c2}$ are the left and right coercive fields, respectively \cite{ref103,ref307}. The variation of H$_{EB}$ calculated from the asymmetric shift of the $M$-$H$ hysteresis loops was inserted in Table \ref{Tab1}. Generally, the EB effect is observed when a system is cooled down in an external magnetic field through the N\'eel temperature (T$_N$). Notably, the BiFeO$_3$ ceramic system exhibited the EB effect without any quintessential method of inducing unidirectional anisotropy \cite{ref311} during the magnetic field annealing process through T$_N$ \cite{ref310}. This ceramic system also showed the EB effect without using any alloy layers \cite{ref312} at RT. The coupling strength of an exchange bias system is indicated by the EB fields. The $H_{EB}$ values inserted in Table \ref{Tab1} are observed without applying any cooling magnetic field and therefore the biasing strength is weak and random. The effect of temperature and cooling magnetic fields on EB effect of this multiferroic system is elaborately described in Ref. \cite{ref101}. Previously, EB effect has been observed in various bulk materials, however, this effect in most cases was limited to far below RT ($<$ 100 K) \cite{ref313, ref314, ref314n} making the systems less lucrative for applications. Therefore, the observation of EB in this co-doped BFO multiferroics up to RT, albeit small, is promising from the perspective of practical applications. 

\begin{figure}[!hh]
	\centering
	\includegraphics[width=9cm]{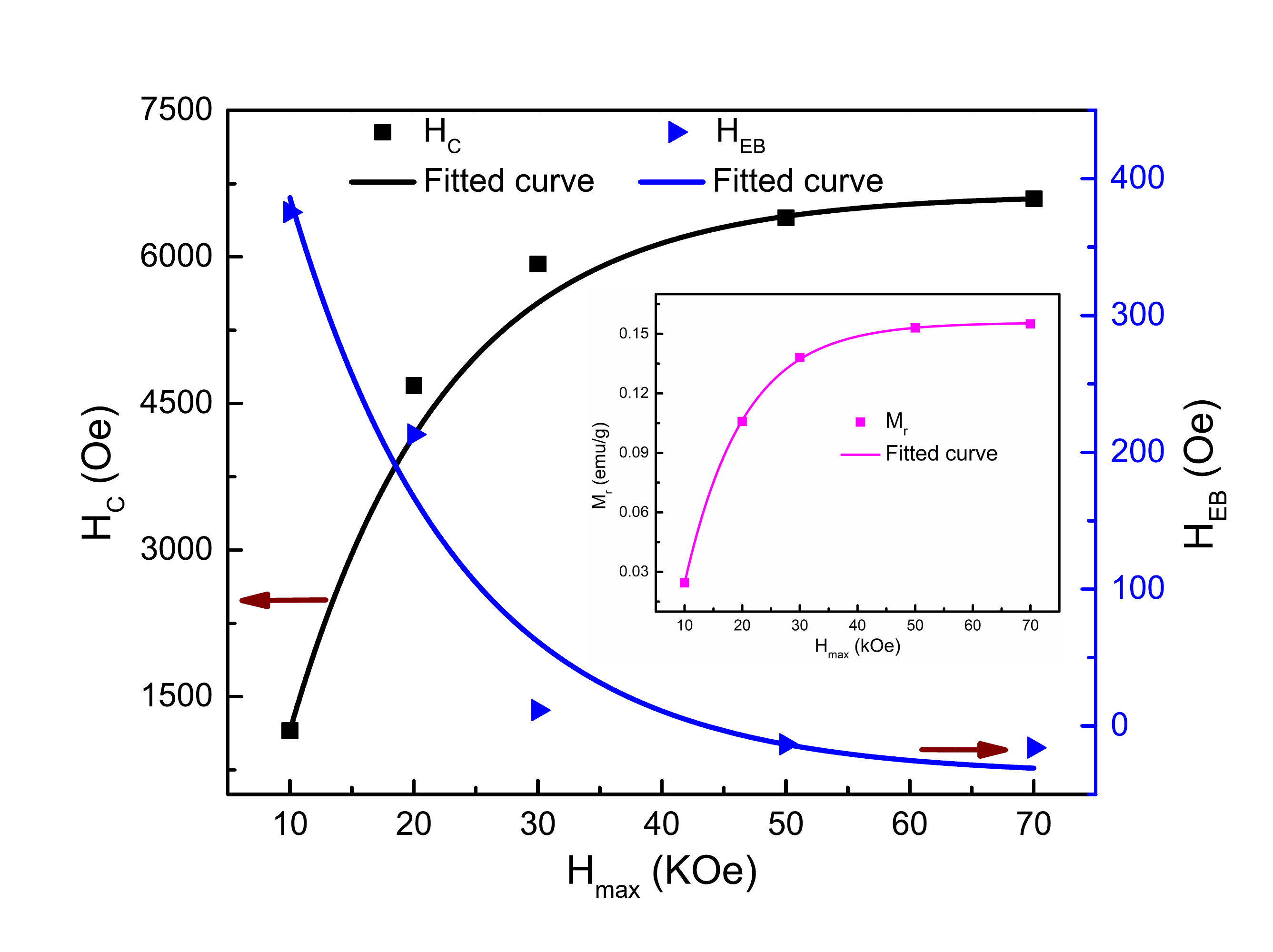}
	\caption{The variation of H$_c$ and H$_{EB}$ of 10\% Gd and Ti co-doped Bi$_{0.9}$Gd$_{0.1}$Fe$_{0.9}$Ti$_{0.1}$O$_3$ material as a function of maximum applied magnetic fields. The inset shows the variation of M$_r$ also as a function of H$_{max}$.} \label{fig3}
\end{figure}

Notably, the $M-H$ hysteresis loops of this multiferroic ceramic at RT demonstrate the unsaturated magnetization behavior even by applying a high magnetic field of up to $\pm$50 kOe. The cation doped multiferroic materials which involve mixed magnetic ordering with large anisotropy do not show a saturating trend even for H $>$ 50 kOe. Thus, the proper choice of maximum field applied for recording a magnetic hysteresis loop, H$_{max}$ is crucial for investigating the magnetization parameters, in particular to investigate the EB effect at RT. Therefore, we have carried out $M-H$ hysteresis loops at different maximum applied magnetic fields (H$_{max}$) for 10\% Gd and Ti co-doped Bi$_{0.9}$Gd$_{0.1}$Fe$_{0.9}$Ti$_{0.1}$O$_3$ sample as showed in Fig.~\ref{fig2}. This particular composition was chosen due to its improved structural properties with significantly reduced impurities. The magnetization parameters were calculated from the respective M-H loops (Fig.~\ref{fig2}). The influence of measuring magnetic fields on the H$_c$ and H$_{EB}$ are shown in Fig. \ref{fig3}. The inset of Fig. \ref{fig3} shows the variation of M$_r$ with H$_{max}$. The H$_c$ and M$_r$ were found to increase with H$_{max}$ while recording a hysteresis loop. The plot of H$_c$ versus H$_{max}$ of Fig.~\ref{fig3} showed that initially when the H$_{max}$ was moderate, the H$_c$ was lower and its variation was linear. On the contrary, H$_c$ increased significantly in a non-linear fashion when H$_{max}$ was increased. In this way, the variation of H$_c$ as a function of H$_{max}$ can be analyzed by dividing the plot of Fig.~\ref{fig3} into two regions, one below 20 kOe (low value of H$_{max}$) and another above 20 kOe (high value of H$_{max}$). Below 20 kOe, the response of H$_c$ to the H$_{max}$ is almost linear. Notably, the unsaturated hysteresis loops of Fig.~\ref{fig2} and large coercive fields (Fig.~\ref{fig3}) extracted from these hysteresis loops confirmed the strong magnetic anisotropy of the synthesized materials. The coercivity is the measure of the magnetic field strength required for overcoming the magnetic anisotropy energy barrier to flip the magnetization. The linear response of the curve of Fig.~\ref{fig3} suggested that magnetic anisotropy is low enough within this low field region and hence H$_c$ was also low. As the H$_{max}$ was increased above 20 kOe, it provided energy to the magnetization which enabled it to cross over the high magnetic anisotropy energy. Therefore, the value of H$_c$ was increased and the response of the H$_c$ to H$_{max}$ became non-linear. At higher H$_{max}$ ($>$50 kOe), the H$_c$ values exhibited an almost saturating trend since the H$_{max}$ values in this range were practically sufficient to overcome the maximum magnetic anisotropy energy of the samples.  

Notably, the H$_{EB}$ were found to decrease with H$_{max}$. Notably, the H$_c$, M$_r$ and H$_{EB}$ values are almost stabilized (considering experimental values) at higher H$_{max}$  i.e. at $>$30 kOe. Moreover, at RT, we observed a transition from negative to positive exchange bias with decreasing H$_{max}$ to record the respective hysteresis loops. A similar transition was observed in La$_{0.5}$Sr$_{0.5}$Mn$_{0.8}$Co$_{0.2}$O$_{3}$ ceramics at temperature much lower than RT \cite{ref314n}.
\begin{figure}[!hh]
	\centering
	\includegraphics[width=8.5cm]{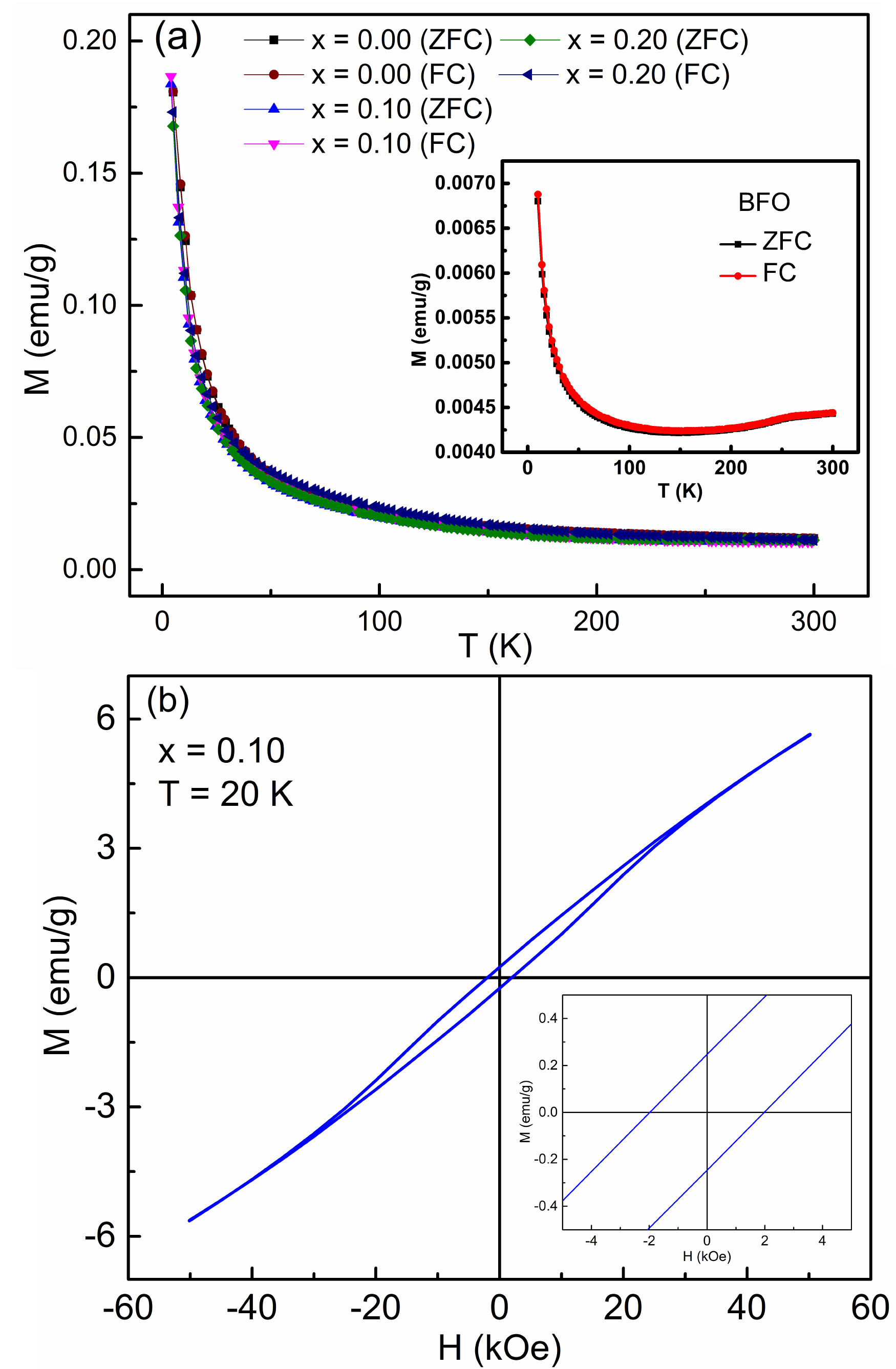}
	\caption{The temperature dependence of magnetization ($M-T$ curves) of Gd doped and Gd-Ti co-doped Bi$_{0.9}$Gd$_{0.1}$Fe$_{1-x}$Ti$_{x}$O$_3$ (x = 0.00, 0.10 and 0.20) materials measured in ZFC and FC processes in the presence of 500 Oe applied magnetic fields. The inset of figure (a) shows ZFC and FC $M-T$ curves of undoped BFO. (b) The M-H hysteresis loop of sample x = 0.10 at 20 K. The inset of (b) shows the enlarged view of this hysteresis loop.} \label{fig4}
\end{figure}

We have also carried out the temperature dependent ZFC and FC magnetization measurements of undoped, Gd doped and Gd-Ti co-doped BFO bulk samples in the presence of 500 Oe applied magnetic field. The $M-T$ curves measured in ZFC and FC modes are shown in Fig.~\ref{fig4} for Gd doped and Gd-Ti co-doped samples. The inset of Fig.~\ref{fig4} (a) demonstrate the $M-T$ curves of the undoped BFO. Initially, cooling was done from 300 K to the minimum obtainable temperature in the ZFC process. In this case, data were collected in the presence of the applied field from the sample while heating. On the contrary, in the FC mode, data collection was performed while the sample was being cooled in the presence of the magnetic field \cite{ref201}.

An anomaly was observed in the ZFC and FC curves of undoped bulk BFO ceramics near 264K, shown in the inset of Fig. \ref{fig4} (a). Later on, the substitution of Gd and co-substitution of Gd and Ti in BFO \cite{ref74, ref306} caused the disappearance of this anomaly. In Ref. \cite{ref74}, it was anticipated that this anomaly originates from domain wall pinning effects due to random distribution of oxygen vacancies. With the decrease in temperature from 300 K to 150 K the magnetization in bulk BFO decreases notably suggesting the AFM nature of the compounds up to 150 K below which (especially below 50K) an abrupt increase of magnetization was observed. Moreover, both ZFC and FC magnetization curves overlapped in the bulk undoped BFO which also suggests their dominating AFM nature as was also observed from the $M-H$ hysteresis loops (Fig. \ref{fig1}). In Gd doped as well as Gd and Ti co-doped BFO ceramics, the ZFC and FC curves exhibit the highest magnetization at the lowest temperature and then there is gradual drop in magnetization with the rise in temperature towards room temperature. The decrease in magnetization is due to the randomization of magnetic spins with the increase in temperature.

%The origin of this sort of anomaly is not fully comprehended. However, Rao \textit{et al.}~\cite{ref74} made an anticipation that domain wall pinning effects, which result from random distribution of oxygen vacancies, might be the underlying cause of such anomalous behavior.
The observed increment in magnetization at lower temperature indicates the interacting magnetic moments where they respond upon the application of low magnetic field (500 Oe). It should be noted that in BFO ceramics the very basis of the canted spin structure is AFM. Therefore these canted spin structures always tend to respond at low temperature, thereby showing an increase in the magnetizations while the temperature is decreased. The ZFC and FC curves for the Gd doped as well as Gd and Ti co-doped BiFeO$_3$ ceramics were found to overlap and 
have not shown any bifurcation \cite{ref206}. 

%This implies that the magnetic response of this material in the presence of an applied magnetic field is dominated by the strong paramagnetic behaviour of Gd$^{3+}$ ion. 

The ZFC and FC magnetization values of Gd doped and Gd-Ti co-doped BFO samples increased with decreasing temperature up to 5 K and also yielded magnetization values much larger than that measured for the undoped BFO indicating the presence of weak FM ordering \cite{ref101, ref306}. To further confirm the weak ferromagnetism at low temperature, an $M-H$ hysteresis loop was carried out at 20 K for 10\% Gd and Ti co-doped Bi$_{0.9}$Gd$_{0.1}$Fe$_{0.9}$Ti$_{0.1}$O$_3$ sample as shown in Fig. \ref{fig4} (b). The inset of Fig. \ref{fig4} (b) shows an enlarged view of the M-H hysteresis loop carried out at 20 K. A tiny loop at the center of the hysteresis with a coercivity of 1979 Oe and remanent magnetization of 0.247 emu/g was observed at 20 K. The coercivity was reduced anomalously at 20 K compared to that at 300 K due to a competition between the magnetic anisotropy and the magnetoelectric coupling of this material as explained in details in Ref. \cite{ref101}.

\subsection{Electrical measurements} \label{IIIB}

\begin{figure}[!t]
	\centering
	\includegraphics[width=8cm]{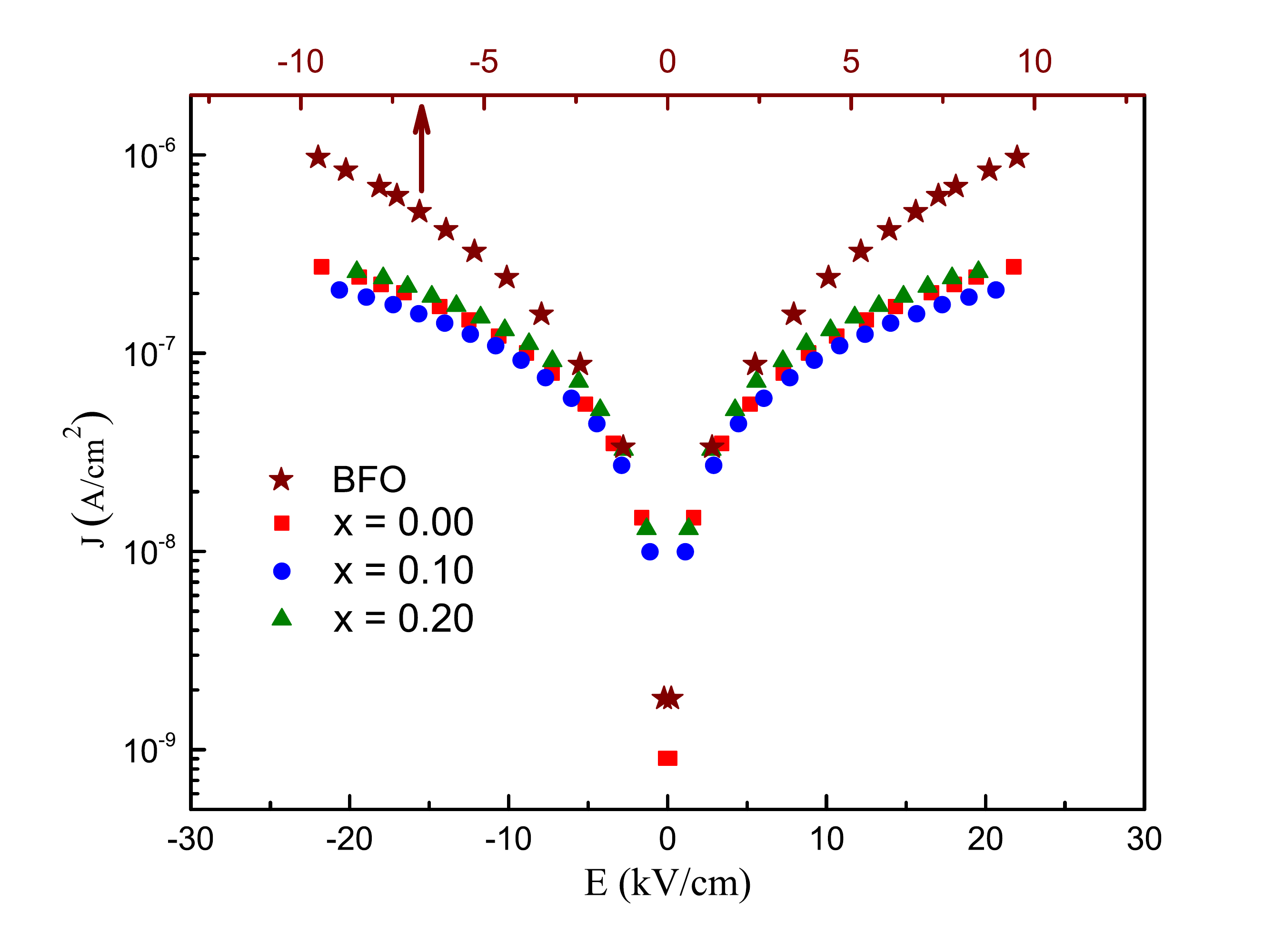}
	\caption{Leakage current density of undoped, Gd doped and Gd-Ti co-doped Bi$_{0.9}$Gd$_{0.1}$Fe$_{1-x}$Ti$_{x}$O$_3$ (x = 0.00, 0.10 and 0.20) materials as a function of applied electric fields. The upper x-axis values correspond to bulk BFO.} \label{fig5}
\end{figure}
To compare the leaky behavior of undoped, Gd doped and Gd-Ti co-doped BFO ceramics, leakage current density, J versus electric field, E measurements were performed. Figure \ref{fig5} shows that the leakage current density of 10\% Gd and Ti co-doped  Bi$_{0.9}$Gd$_{0.1}$Fe$_{0.9}$Ti$_{0.1}$O$_3$ bulk materials is smaller than that of other compositions. Particularly, the leakage current density of this co-doped ceramic is much smaller than that of undoped BFO. The high leakage current of bulk material is predominantly connected with impurity phases and oxygen vacancies \cite{ref501, ref502}. As previously demonstrated in Fig. \ref{fig01}, for the 10\% Gd and Ti co-doped  Bi$_{0.9}$Gd$_{0.1}$Fe$_{0.9}$Ti$_{0.1}$O$_3$, impurity phases were inhibited significantly. Again, from Fig. \ref{fig02}, we observed a noticeable reduction in oxygen vacancies. Therefore, it is expected that due to the substitution of Gd and Ti in BiFeO$_3$,  the oxygen vacancies induced mainly due to the volatilization of Bi$^{3+}$ ion in the lattice were suppressed and consequently the leakage current density was reduced. 

\begin{figure}[!hh]
	\centering
	\includegraphics[width=9cm]{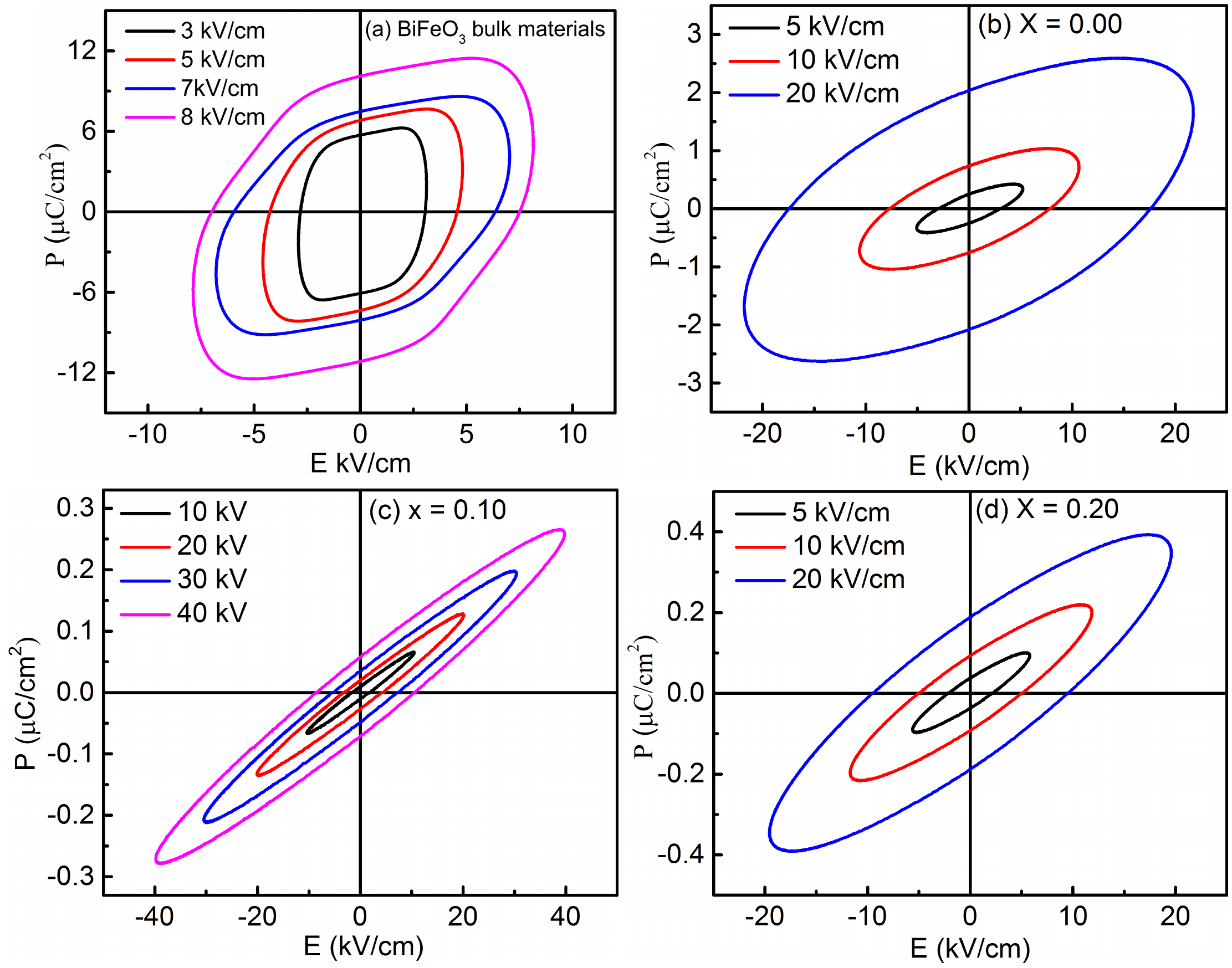}
	\caption{The P-E hysteresis loops of (a) undoped BFO, (b) Gd doped Bi$_{0.9}$Gd$_{0.1}$Fe$_{1-x}$Ti$_x$O$_3$ (x = 0.00), (c) Gd-Ti co-doped Bi$_{0.9}$Gd$_{0.1}$Fe$_{1-x}$Ti$_x$O$_3$ (x = 0.10), and (d) Gd-Ti co-doped Bi$_{0.9}$Gd$_{0.1}$Fe$_{1-x}$Ti$_x$O$_3$ (x = 0.20) materials.} \label{fig6}
\end{figure}

Figure \ref{fig6} (a-d) shows the ferroelectric polarization hysteresis loops (P-E loops) of undoped, Gd doped and Gd-Ti co-doped BFO ceramic systems which were carried out by varying E-fields. Due to the different breakdown fields for different samples, the applied electric fields were varied. The continuous increment of electric field resulted in the increased remanent polarization while the driving frequency was maintained at 50 Hz. This is due to the fact that larger electric field provides higher level  of driving power responsible for reversal of ferroelectric domains \cite{ref503}. It is expected that in undoped BFO, freely movable charges appear to contribute significantly to the electrical hysteresis loop.  Therefore, the undoped BFO material exhibits a round shaped P-E loop as shown in Fig. \ref{fig6} (a) due to its high leakage current as was evidenced in Fig. \ref{fig5}. The substitution of Gd in place of Bi in  BiFeO$_3$ i.e in sample x = 0.00 (Fig. \ref{fig6} (b)) reduces the contribution to the polarization from the freely movable charges, as evidenced by the less rounded features of the loops. The leakage current density was also found to decrease upon the substitution of Gd in place of Bi in BiFeO$_3$ compared to that of undoped BFO. The rounded shape P-E loops for undoped and 10\% Gd doped BFO with large P$_r$ indicate that the electrical leakage is severe in these materials.

 In the case of 10\% Gd and Ti co-doped  Bi$_{0.9}$Gd$_{0.1}$Fe$_{0.9}$Ti$_{0.1}$O$_3$ i.e. for x = 0.10 composition, Fig. \ref{fig6} (c), the P-E loops became more and more typical which is associated with their reduced leakage current density \cite{ref504} compared to that of all other compositions. For a further increment of Ti concentration to 20\% i.e. for sample x = 0.20, Fig. \ref{fig6} (d), the shape of the loop remains elliptical, however, the remanent polarization is enhanced again probably due to the contribution of the increased freely movable charges. Hence, the electrical measurements clearly show an improved ferroelectric behavior of 10\% Gd and Ti co-doped BFO material due to its reduced leakage current density.
 
 %\textcolor {red} {Notably, in comparison with the undoped and 10\% Gd doped BFO, we did not observe rounded P-E loops for Gd-Ti co-doped BFO samples which meant there was no contribution from the freely movable charges in the P-E loops of co-doped ceramics.}

The comparison of the P-E loops between undoped and doped BiFeO$_3$ materials, (Fig. \ref{fig6} (a) and (b-d)) also demonstrates that the breakdown voltage is significantly increased due to substitution of Gd and co-substitution of Gd and Ti. By comparing P-E loops of Fig. \ref{fig6} and  dielectric constant observed in Ref. \cite{ref6}, it is obvious that the remanent polarization and dielectric constant follow exactly the same trend. This similarity is expected as the polarization is related to electric field through the dielectric constant.

\section{Conclusions} \label{II}

In this investigation, field dependent magnetization measurements were carried out in undoped, Gd doped and Gd-Ti co-doped BiFeO$_3$ multiferroic ceramics. For 10\% Gd and Ti co-doped  Bi$_{0.9}$Gd$_{0.1}$Fe$_{0.9}$Ti$_{0.1}$O$_3$ bulk material, the remanent magnetization and coercive field are found to be maximum compared to that of other compositions. The influence of maximum applied magnetic fields on H$_c$, M$_r$ and H$_{EB}$ was found to be stabilized at higher H$_{max}$ ($>$30 kOe). This multiferroic ceramic material exhibited EB effect notably at RT which indicates the coexistence of their AFM and FM orderings. We also obtained a transition from negative to positive EB with decreasing H$_{max}$ in this material, that again at RT, which could lead to various exciting applications. The ZFC and FC magnetization curves of undoped bulk BiFeO$_3$ ceramics show an anomaly near 264 K which was found to disappear upon the substitution of Gd and also co-substitution of Gd and Ti in BiFeO$_3$. Due to the substitution of rare-earth Gd and transition metal Ti in place of Bi and Fe in BiFeO$_3$, respectively, the leakage current density was reduced. This reduction in leakage current densities might be due to reduction of oxygen vacancy related defects which were analysed from the XPS spectrum. The value of the leakage current density was minimum for 10\% Gd and Ti co-doped  Bi$_{0.9}$Gd$_{0.1}$Fe$_{0.9}$Ti$_{0.1}$O$_3$  bulk material and therefore an improved ferroelectric behavior of this composition was observed. The improved multiferroic properties obtained by 10\% Gd and Ti co-doping in BiFeO$_3$ is promising for novel multifunctional device applications. We may further conclude that an appropriate choice of co-doping elements and fine composition adjustment are keys to optimize the multiferroic properties of BiFeO$_3$ ceramics.

%It is expected that the observed dielectric anomaly of 10\% Gd and Ti co-doped  Bi$_{0.9}$Gd$_{0.1}$Fe$_{0.9}$Ti$_{0.1}$O$_3$ material around  N\'eel temperature may be a consequence of magnetoelectric coupling.

\section*{ACKNOWLEDGMENTS}
This work was supported by Ministry of Science and Technology, Government of Bangladesh, Grant No.:39.009.002.01.00.053.2014-2015/PHY’S-273/ (26.01.2015) and The World Academy of Sciences (TWAS), Ref.:14-066 RG/PHYS/AS-I; UNESCO FR: 324028567. The Institute for Molecular Science (IMS), supported by Nanotechnology Platform Program (Molecule and Material Synthesis) of MEXT, Japan for providing SQUID facilities.

\end{document}